\documentstyle[12pt,epsfig]{article}

\newcommand{\beq}{\begin{equation}}
\newcommand{\eeq}{\end{equation}}
\newcommand{\beqa}{\begin{eqnarray}}
\newcommand{\eeqa}{\end{eqnarray}}

\newcommand{\labell}[1]{\label{#1}} 
\newcommand{\reef}[1]{(\ref{#1})}

\def\p{\partial}

\def\ie{{\it i.e.,}\ }
\def\eg{{\it e.g.,}\ }

\let\bar=\overbar

\def\ie{{\it i.e.}}
\def\eg{{\it e.g.}}

\def\Dslash{\not{\hbox{\kern-4pt $D$}}}
\def\dslash{\not{\hbox{\kern-2pt $\del$}}}

\def\p4n{{\mbox{\scriptsize (4+N)}}}
\def\p4{{\mbox{\scriptsize (4)}}}

\def\msb{{\bar{\ssstyle M \kern -1pt S}}}

\begin{document}

\setlength{\unitlength}{1mm}

\thispagestyle{empty}
\rightline{\small hep-th/0111178 \hfill}
\vspace*{2cm}

\begin{center}
{\bf \LARGE Superstars and Giant Gravitons}\\[.5em]
{\bf \LARGE  in M-Theory}
\vspace*{1cm}

{\bf Fr\'ed\'eric Leblond}\footnote{E-mail: {\tt fleblond@hep.physics.mcgill.ca}},
{\bf Robert~C.~Myers}\footnote{E-mail: {\tt rcm@hep.physics.mcgill.ca}}
{\bf and
David~C.~Page}\footnote{E-mail: {\tt d.c.page@durham.ac.uk}}

\vspace*{0.3cm}
\vspace{0.5cm}
$^{1,2}$Department of Physics, McGill University \\
3600 University Street, Montr\'eal, Qu\'ebec H3A 2T8, Canada\\
\vspace{0.5cm}
$^{2}$Perimeter Institute for Theoretical Physics \\
35 King Street North, Waterloo, Ontario N2J 2W9, Canada \\
\vspace{0.5cm}
$^{2}$Department of Physics, University of Waterloo \\
Waterloo, Ontario N2L 3G1, Canada\\
\vspace{0.5cm}
$^{3}$Centre for Particle Theory \\
Department of Mathematical Sciences \\
University of Durham, Durham DH1 3LE, U.K. \\

\vspace{2cm}
{\bf ABSTRACT}
\end{center}

Following hep--th/0109127, we show that a certain class of BPS naked singularities
(superstars) found in compactifications of M--theory can be interpreted as being composed
of giant gravitons. More specifically, we study superstars which are asymptotically
$AdS_{7}\times S^{4}$ and $AdS_{4}\times S^{7}$ and show that these field configurations
can be interpreted as being sourced by continuous distributions of spherical
M2-- and M5--branes, respectively, which carry internal momenta and have expanded on the
spherical component of the space--time.

\vfill
\setcounter{page}{0}
\setcounter{footnote}{0}
\newpage

\section{Introduction}

In a recent paper\cite{myers1}, convincing evidence was provided showing
that certain supersymmetric naked singularities, appearing in type IIB supergravity
compactified on $AdS_{5}\times S^{5}$, can be interpreted in terms of distributions
of giant gravitons\cite{toumbas}. These `superstar' solutions correspond to the supersymmetric
limit of a certain family of black holes. The latter were originally found when considering
a consistent truncation of the type IIB supergravity theory dimensionally reduced
to $AdS_{5}$ where they appear as charged black hole solutions. Once lifted back to the
full ten--dimensional theory, the solutions carry internal momentum along the three
commuting Killing angles on the five--sphere, and the supersymmetric limit still leaves
a naked singularity. However, within the context of ten--dimensional type IIB superstring
theory, there is a physical interpretation in which the singularities are generated
by a distribution of giant gravitons, \ie, an ensemble of spherical D3--branes which
carry internal angular momentum and have expanded on the five--sphere\cite{toumbas}.
This result was determined by examining the dipole field excited in the Ramond--Ramond
five--form near the singularity of the supergravity solutions.

In this note, we study the analogous superstar solutions in M--theory compactified on
$AdS_{7}\times S^{4}$ and $AdS_{4}\times S^{7}$. We show that these eleven--dimensional
supergravity solutions can also be interpreted as being sourced by distributions of giant
gravitons. Hence in this case, the constituent degrees of freedom
are M2 and M5--branes, respectively. In the second part of this note, we
study the behavior of test--brane probes in the background geometry of the
M--theory superstars. By considering giant graviton probes, we are able to
confirm the expansion of these configurations.
We conclude with a brief discussion of our results. We consider dual giant
graviton\cite{myers2,hashi} probes of superstars in an appendix.

\section{Superstars in $AdS_{7}\times S^{4}$}
\label{section1}
Eleven--dimensional supergravity appears as a particular low energy limit of
M--theory\cite{witten}. The bosonic sector of this theory is composed
of the graviton and the four--form field strength $F_{(4)}$. The latter allows
for a spontaneous compactification\cite{freund} of the theory on $AdS_{7}\times S^{4}$.
Further in this background, there is a consistent truncation of the full theory
to ${\cal N}=4$ gauged $SO(5)$ supergravity in seven dimensions\cite{vaman1,vaman2}.
In this note, we focus on solutions of a specific ${\cal N}=2$ truncation of this theory
where only the following seven--dimensional fields are retained:
the metric, two scalars labelled by $X_{i}$ ($i=1,2$) and two one--form gauge fields
$A^{i}_{(1)}$ associated with the $U(1)^2$ Cartan subgroup of $SO(5)$
(see, \eg, refs. \cite{duff1,duff2} for details). The correct Kaluza--Klein
$S^{4}$ reduction ansatz, which is going to be used to lift seven--dimensional solutions
to solutions of the full eleven--dimensional supergravity equations of motion,
is\cite{duff1,duff2}
\beq
\labell{metric1}
ds_{11}^{2}=\tilde{\Delta}^{1/3}ds_{7}^{2} + \frac{L^2}{\tilde{\Delta}^{2/3}}
\left( \frac{1}{X_{0}} d\mu_{0}^{2} + \sum_{i=1}^{2}\frac{1}{X_{i}}
\left[ d\mu_{i}^{2} + \mu_{i}^{2} \left( d\phi_{i} + A^{i}_{(1)}/L \right)^{2} \right] \right)
\eeq
for the metric, and
\begin{eqnarray}
\labell{3form1}
*F_{(4)} = -\frac{2}{L} \sum_{\alpha=0}^{2} \left( X_{\alpha}^{2}\mu_{\alpha}^{2}
- \tilde{\Delta}X_{\alpha} \right)
\epsilon_{(7)} - \frac{1}{L} \tilde{\Delta}X_{0}\epsilon_{(7)}
- \frac{L}{2}\sum_{\alpha=0}^{2}\frac{1}{X_{\alpha}}
\bar{*}dX_{\alpha}\wedge d\mu_{\alpha}^{2} \nonumber \\
- \frac{L^2}{2}\sum_{i=1}^{2}\frac{1}{X_{i}^{2}}d\mu_{i}^{2}\wedge
\left( d\phi_{i} + A^{i}_{(1)}/L \right) \wedge \bar{*} F^{i}_{(2)},\qquad
\end{eqnarray}
for the field strength of the supergravity three--form. The internal four--sphere
is parameterized by four angular variables: the $\phi_{i}$'s are azimuthal (Killing) angles
and the other angles are defined through the direction--cosines
\beq
\mu_{0} = \sin \theta_{1} \sin \theta_{2} \; \; \; \; \mu_{1} = \cos \theta_{1} \; \; \; \;
\mu_{2} = \sin \theta_{1} \cos \theta_{2} \; ,
\eeq
such that $\mu_{0}^{2}+\mu_{1}^{2}+\mu_{2}^{2}=1$. The $\bar{*}$ denotes the Hodge dual
with respect to the seven--dimensional metric $ds_{7}^2$. Finally, $L$ is the (asymptotic)
radius of curvature of the four--sphere, $X_{0}\equiv (X_{1}X_{2})^{-1}$ and
$\tilde{\Delta}=\sum_{\alpha=0}^{2}X_{\alpha}\mu_{\alpha}^{2}$.

The seven--dimensional ${\cal N}=2$ supergravity admits doubly--charged $AdS$ black hole
solutions of the form\cite{gubser},
\beq
\labell{ads7m}
ds^{2}_{7} = -\frac{f}{\left(H_{1}H_{2}\right)^{4/5}}dt^{2} + \left( H_{1}H_{2} \right)^{1/5}
\left( \frac{1}{f}dr^{2} + r^{2} d\Omega_{5}^{2} \right),
\eeq
\beq
\labell{ads7g}
A_{(1)}^{i} = \left(\frac{1}{H_{i}} -1 \right) dt \;\;\;\;\;\;\;\;\; i=1,2 \;,
\eeq
\beq
\labell{ads7s}
X_{i} = \frac{\left(H_{1}H_{2}\right)^{2/5}}{H_{i}},
\eeq
where we have introduced
\beq
f = 1 - \frac{\mu}{r^{4}} + \frac{r^2}{4L^2}H_{1}H_{2},
\eeq
\beq
H_{i} = 1 + \frac{q_{i}}{r^{4}},
\eeq
with $\mu$ a mass parameter and the $q_{i}$'s related to the physical charges of the black
hole. We will use the following expression for the metric on the five--dimensional unit sphere,
\beq
d\Omega_{5}^{2} = d\alpha_{1}^{2} + \sin^{2} \alpha_{1} \left( d\alpha_{2}^{2}
+\sin^{2}\alpha_{2}
\left[ d\alpha_{3}^{2} + \sin^{2} \alpha_{3} \left( d\alpha_{4}^{4} +
\sin^{2} \alpha_{4} d\alpha_{5}^{2} \right) \right] \right).
\eeq
The mass of this $AdS_{7}$ black hole is\cite{adsbh}
\beq
\labell{adsmass}
M=\frac{\pi^{2}}{4G_{7}}\left( \frac{5}{4}\mu + \sum_{i} q_{i} \right),
\eeq
a quantity that will later be compared with the energy of the suggested
ensemble of source branes.

The horizon structure of both the metric eq.~(\ref{ads7m}) and its lift to eleven--dimensional
supergravity using ansatz eq.~(\ref{metric1}) are essentially the same as can be seen by
analysing the $(rr)$ component of the corresponding metrics. The supersymmetric limit
corresponds to $\mu=0$ where the event horizon vanishes to reveal
a naked singularity. Hence we denote these BPS configurations as `superstars.'
For $\mu=0$, the entropy vanishes which means that the system should have a very simple
physical interpretation in terms
of a single arrangement of fundamental degrees of freedom. It is natural to conjecture
that these degrees of freedom are giant gravitons, just as in the type IIB case\cite{myers1}.
The $\mu \neq 0$ black holes of finite entropy are likely still related to a
similar ensemble of expanded branes but non--trivial interactions in the non--extremal case
complicate the analysis. Hence we restrict ourselves to studying the BPS objects.

Following ref.~\cite{myers1} then, we make the conjecture that the superstars can be given
a reasonable physical interpretation as the external fields around an ensemble of giant
gravitons. The first evidence of this conjecture is simply that the off--diagonal
form of the eleven--dimensional metric \reef{metric1} shows that the superstars carry
internal momentum along the $\phi_i$ directions, just as giant gravitons would.
However if the $q_{i}$'s are to be macroscopic distance scales (much larger than the eleven--dimensional
Planck scale), then
the mass of the superstar must be significantly larger than the energy of an
individual giant graviton. This fact suggests that superstars
must be composed of a collection of these fundamental objects.

For the detailed analysis, we begin for simplicity by considering superstars that
are singly charged: $q_{1}\neq 0$ and $q_{2}=0$. Since the source M2--branes must carry
internal momentum along $\phi_{1}$, they should
span the two--sphere parameterized by $\theta_{2}$ and $\phi_{2}$. The collection
of giant gravitons each with varying momenta should be distributed
along the $\theta_{1}$--direction.
Using a test--brane analysis\cite{toumbas}, one finds that for a given
internal momentum $P_{\phi_{1}}$ (the canonical conjugate to $\phi_{1}$),
a spherical M2--brane configuration on $AdS_{7}\times S^{4}$ has two minima, one
at $\sin \theta_{1} = 0$ and the other at $\sin \theta_{1} = P_{\phi_{1}}/N$, where $N$
is the number of four--form flux quanta on the four--sphere. The former corresponds
to a point--like graviton with angular momentum $P_{\phi_{1}}$, and the latter to the
giant graviton, with
the same angular momentum, that has expanded to a fixed size on a two--dimensional
submanifold of the $S^{4}$. The point--like and giant gravitons preserve the same number of
supersymmetries\cite{myers2,hashi} and have the same energy\cite{toumbas}:
\beq
\labell{energyGG}
E = \frac{P_{\phi_{1}}}{L} = \frac{N}{L}\sin \theta_{1}.
\eeq
It is natural to assume that the superstar source must correspond to the expanded
giant gravitons as the singularity in the metric extends over the entire four--sphere
at $r=0$.

A M2--brane is electrically charged with
respect to the supergravity three--form potential. However,
the spherical M2--branes will carry no net charge, but they will locally
excite this field. If one considers a small (seven--dimensional) surface that encloses
a portion of the sphere, one will find a net flux proportional to the
number of M2--branes enclosed. In fact,
\beq
\labell{charge}
\int_{M_{7}} *F_{4}  = 16 \pi G_{11} T_{M2} n_{1},
\eeq
where $G_{11}$ is the eleven--dimensional gravitational constant, $T_{M2}$ the tension of a
M2--brane and $n_{1}$ the total number of giant gravitons enclosed. We are using the
following conventions,
\beq
\labell{g11}
G_{11} = 16 \pi^{7} l_{P}^{9},
\qquad
G_{7} = \frac{G_{11}}{V_{S^{4}}} = \frac{3}{8\pi^{2}L^{4}}G_{11},
\qquad
T_{M2} = \frac{1}{4\pi^{2}l_{P}^{3}},
\eeq
where $l_{P}$ is the fundamental eleven--dimensional Planck length. Given the orientation
of the giant gravitons described above, the surface
$M_{7}$ is a manifold in the subspace spanned by $\lbrace \theta_{1},\phi_1,\alpha_i\rbrace$.
If we wish to consider the entire distribution, we naturally fix $r$ and integrate over the remaining
angles\cite{myers1}. The $\phi_{1}$ integration is of course trivial because the supergravity
solution is smeared along this direction.
In evaluating the expression
eq.~(\ref{charge}), we find that the only term in eq.~(\ref{3form1}) which gives a
nonvanishing contribution is
\beq
-\frac{L^{2}}{2 X_{1}^{2}} d\mu_{1}^{2} \wedge d\phi_{1} \wedge \bar{*} F_{(2)},
\eeq
where
\beq
\left[\bar{*}F_{(2)}\right]_{\alpha_{1}\alpha_{2}\alpha_{3}\alpha_{4}\alpha_{5}} =
-\frac{4q_{1}}{H_{1}^{6/5}} \sin^{4} \alpha_{1} \sin^{3} \alpha_{2}
\sin^{2} \alpha_{3} \sin \alpha_{4}.
\eeq
Hence the total number of giant gravitons $n_{1}$ is found to be
\beq
\labell{number}
n_{1} = \frac{ q_{1}L^{2}}{4\pi G_{11} T_{M2}} \int d\theta_{1} \, d\phi_{1} \,d^5\alpha_{i}
\; \cos \theta_{1} \sin \theta_{1} \sin^{4} \alpha_{1}
\sin^{3} \alpha_{2} \sin^{2} \alpha_{3} \sin \alpha_{4}.
\eeq
Dropping the $\theta_{1}$ integration (along which the giant gravitons are assumed to be distributed) from eq.~(\ref{number})
and integrating over the other angles
we find an expression for the density of giant gravitons as a function of $\theta_{1}$,
\beq
\labell{distrib2}
\frac{dn_{1}}{d\theta_{1}} = \frac{q_{1}N^{2}}{8 L^{4}} \cos \theta_{1} \sin \theta_{1} .
\eeq
Given that we have detected a distribution of `electric dipole' sources for $F_{(4)}$
at the singularity is certainly evidence of the presence of giant gravitons.

Further support for this result comes by considering the energy of the above
distribution\cite{myers1}. The test--brane analysis\cite{toumbas} found that the size
and internal momentum of an individual giant graviton were related by:
$P_{\phi_{1}} = N \sin \theta_{1}.$ Combining this result with eq.~\reef{distrib2},
we find the total angular momentum of the distribution to be
\beq
\labell{totp}
\bar{P}_{\phi_{1}} = N \int_{0}^{\pi/2} d\theta_{1} \sin \theta_{1} \frac{dn_{1}}{d\theta_{1}} = \frac{1}{24}
\frac{q_{1} N^{3}}{L^{4}}.
\eeq
According to eq.~(\ref{energyGG}), the total energy is then
\beq
\labell{totalenergy}
\bar{E} = \frac{1}{24} \frac{q_{1} N^{3}}{L^{5}}.
\eeq
The mass of the superstar \reef{adsmass}, calculated by conventional supergravity
means, is
\beq
\labell{1charge}
M=\frac{\pi^{2}q_{1}}{4G_{7}}.
\eeq
Using eqs.~(\ref{g11}) to write this mass in terms of $N$ and $L$, one
shows that it agrees exactly with the total energy eq.~(\ref{totalenergy})
of a distribution of giant gravitons.
As it stands this result is somewhat of a curiousity. It is not clear why the result
for the size of the giant graviton derived from a test--brane propagating in the
$AdS_7\times S^4$ background should apply here. However, we will find supporting
evidence for this fact from our probe analysis in section 4.

It is straightforward to generalize the calculation to the case of superstars with two
non--vanishing charges: $q_{1}\neq 0$ and $q_{2}\neq 0$. We consider the following embedding
of $S^{4}$ in ${\bf R}^{5}$ with coordinates $x^{0,1,2,3,4}$,
\beq
x^{0} = L \mu_{0}, \;\;\; x^{2i-1} = L \mu_{i} \cos \phi_{i}, \;\;\;
x^{2i}=L\mu_{i}\sin \phi_{i},
\eeq
where $i=1,2$. Consequently, a giant graviton moving along $\phi_{i}$ has a radius
\beq
\rho_{i} = L \sqrt{1-\mu_{i}^{2}}.
\eeq
In analogy with the single charge calculation, the density of gravitons of a certain radius
involves the ${*}F^{(4)}$,
\beq
\left[{*}F^{(4)}\right]_{\rho_{i}\phi_{i}\alpha_{1}\alpha_{2}\alpha_{3}\alpha_{4}\alpha_{5}} =
\frac{d\mu_{i}^{2}}{d\rho_{i}} \left[{*}F^{(4)}\right]_{\mu_{i}^{2}
\phi_{i}\alpha_{1}\alpha_{2}\alpha_{3}\alpha_{4}\alpha_{5}}
= 4q_{i}\rho_{i} \sin^{4} \alpha_{1}
\sin^{3} \alpha_{2}\sin^{2} \alpha_{3}\sin \alpha_{4}.
\eeq
The corresponding density of giant gravitons for each direction is then
\beq
\frac{dn_{i}}{d\rho_{i}} = \frac{q_{i}N^{2}\rho_{i}}{L^{6}}.
\eeq
Treating this as two independent distributions,
it is found that the total angular momentum carried by each set of giant gravitons is
\beq
\bar{P}_{\phi_{i}} = \frac{1}{24} \frac{q_{i} N^{3}}{L^{4}}.
\eeq
These results match the total angular momentum calculated for the superstar solution,
and we have complete agreement between the BPS mass of the superstar (\ref{adsmass})
and the total energy of the giant gravitons, $\bar{E}=\sum \bar{P}_{\phi_{i}}/L$.

\section{Superstars in $AdS_{4}\times S^{7}$}

In this section we consider superstars which are asymptotically $AdS_{4}\times S^{7}$. Their
description is hypothesized to be in terms
of M5--branes carrying internal momentum that have expanded on the $S^{7}$.
Eleven--dimensional supergravity admits a spontaneous compactification on $AdS_{4}\times S^{7}$.
The Kaluza--Klein reduction of
this theory on $S^{7}$ leads to ${\cal N}=8$ gauged $SO(8)$ supergravity in four dimensions.
We focus on a specific ${\cal N}=2$ truncation where only the following fields are retained:
the metric, four scalars labelled by $X_{i}$ ($i=1,2,3,4$) ($X_{1}X_{2}X_{3}X_{4}=1$)
and four one--form gauge fields $A^{i}_{(1)}$
(see, \eg, refs. \cite{duff1,duff2} for details). The correct Kaluza--Klein
$S^{7}$ reduction ansatz, which is going to be used to lift four--dimensional solutions to solutions of
the full eleven--dimensional supergravity equations of motion, is\cite{duff1,duff2}
\beq
\labell{metric2}
ds_{11}^{2}=\tilde{\Delta}^{2/3}ds_{4}^{2} + \frac{L^{2}}{\tilde{\Delta}^{1/3}}
\sum_{i=1}^{4} \frac{1}{X_{i}} \left( d\mu_{i}^{2} + \mu_{i}^{2} \left( d\phi_{i} + A^{i}_{(1)}/L \right)^{2} \right)
\eeq
for the metric, and
\begin{eqnarray}
\labell{3form2}
F_{(4)} = \frac{2}{L} \sum_{i=1}^{4}\left( X_{i}^{2} \mu_{i}^{2} - \tilde{\Delta} X_{i} \right)\epsilon_{(4)}
+ \frac{L}{2}\sum_{i=1}^{4} \frac{1}{X_{i}} \bar{*}dX_{i}\wedge d\mu_{i}^{2} -
\frac{L^{2}}{2} \sum_{i=1}^{4}\frac{1}{X_{i}^{2}} d\mu_{i}^{2} \wedge \left( d\phi_{i} + A^{i}_{(1)}/L \right)
\wedge \bar{*}F^{i}_{(2)}  ,
\end{eqnarray}
for the field strength of the supergravity three--form. The space transverse to $ds_{4}^{2}$ is characterized
by seven angular variables: the $\phi_{i}$'s are azimuthal (Killing) angles and the other angular variables
are defined through the direction--cosines
\beq
\mu_{1} = \cos \theta_{1} \; \; \; \mu_{2} = \sin \theta_{1} \cos \theta_{2} \; \; \; \mu_{3}
= \sin \theta_{1} \sin \theta_{2} \sin \theta_{3}
 \; \; \; \mu_{4} = \sin \theta_{1} \cos \theta_{2} \cos \theta_{3} ,
\eeq
such that $\mu_{1}^{2}+\mu_{2}^{2}+\mu_{3}^{2}+\mu_{4}^{2}=1$. Finally, $L$ is the (asymptotic)
radius of curvature of the seven--sphere and
$\tilde{\Delta}=\sum_{i=1}^{4}X_{i}\mu_{i}^{2}$.

The corresponding four--dimensional action for the graviton, the scalars $X^{i}$ and the one--form fields
$A_{(1)}^{i}$ admit a quadruply charged $AdS$ black hole solution\cite{duff3,sabra},
\beq
\labell{ads4m}
ds^{2}_{4} = -\frac{f}{\left(H_{1}H_{2}H_{3}H_{4}\right)^{1/2}}dt^{2} + \left( H_{1}H_{2}H_{3}H_{4} \right)^{1/2}
\left( \frac{1}{f}dr^{2} + r^{2} d\Omega_{2}^{2} \right),
\eeq
\beq
\labell{ads4g}
A_{(1)}^{i} = \left( \frac{1}{H_{i}} - 1 \right) dt \;\;\;\;\;\;\;\;\; i=1,2,3,4 \;,
\eeq
\beq
\labell{ads4s}
X_{i} = \frac{\left(H_{1}H_{2}H_{3}H_{4}\right)^{1/4}}{H_{i}},
\eeq
where we have introduced
\beq
f = 1 - \frac{\mu}{r} + \frac{4r^{2}}{L^{2}}H_{1}H_{2}H_{3}H_{4},
\eeq
\beq
H_{i} = 1 + \frac{q_{i}}{r},
\eeq
with $\mu$ a mass parameter and the $q_{i}$'s related to the physical charges of the black hole.
We use the following expression
for the metric on the two--dimensional unit sphere,
\beq
d\Omega_{2}^{2} = d\alpha_{1}^{2} + \sin^{2} \alpha_{1} d\alpha_{2}^{2} .
\eeq
The mass of this $AdS_{4}$ black hole is\cite{adsbh}
\beq
\labell{adsmass2}
M=\frac{1}{4G_{4}}\left( 2\mu + \sum_{i} q_{i} \right),
\eeq
a quantity that is be compared with the energy of the hypothesized
equivalent system composed of spherical M5--branes. Once again, the horizon
structure of the lower dimensional realisation of the black hole and its lift to eleven--dimensional supergravity
using ansatz eq.~(\ref{metric2}) are essentially the same. We assume that
\beq
q_{1}\geq q_{2} \geq q_{3} \geq q_{4}.
\eeq
For $q_{1}\neq 0$ with all other charges vanishing, we find that for $\mu=0$ there is no horizon behind which
the curvature singularity at $r=0$ is hidden. Whenever two charges or more have a finite value,
there is a critical value for the non--extremality parameter, $\mu =\mu_{crit}$, corresponding to
dissapearing horizons. In other words, for $\mu > \mu_{crit}$ the geometry corresponds to a regular
black hole and for $\mu < \mu_{crit}$, it corresponds to a naked singularity \ie,  to the so--called
superstars. In any case, a generic feature is that for $\mu=0$ (BPS limit)
the area of the spherical horizon is zero.

Following the discussion of section \ref{section1}, we consider superstars that
are singly charged ($q_{1}\neq 0$ and $q_{2}=q_{3}=q_{4}=0$) and assume that they are composed of M5--branes
spanning a five--sphere parametrized by $\theta_{2}$, $\theta_{3}$ and $\phi_{2}$, $\phi_{3}$, $\phi_{4}$. The
giant gravitons have non--zero internal momentum along $\phi_{1}$,
are distributed along the $\theta_{1}$--direction and have an energy given by\cite{toumbas}
\beq
\labell{energyGG2}
E = \frac{P_{\phi_{1}}}{L} = \frac{N}{L}\sin^{4} \theta_{1},
\eeq
where $N$ is the number of seven--form (the dual of $F^{(4)}$) flux quanta on the seven--sphere.
A M5--brane is magnetically charged with respect to the supergravity three--form potential.
The spherical M5--branes carry no net charge but nevertheless excite the field locally. If one considers
a small surface enclosing a portion of the sphere, one finds a net flux proportional to the
number of M5--branes enclosed,
\beq
\labell{charge2}
\int_{M_{4}} F_{4}  = 16 \pi G_{11} T_{M5} n_{1},
\eeq
where $G_{11}$ is the eleven--dimensional gravitational constant, $T_{M5}$ is the tension of a
M5--brane and $n_{1}$ is the total number of giant gravitons making the superstar. Given the orientation
of the giant gravitons described above, $M_{4}$ is a manifold in the subspace spanned by
$\lbrace \theta_{1},\phi_1,\alpha_i\rbrace$.
The $\phi_{1}$ integration is trivial because the supergravity solution is smeared along
that direction. We are using the following conventions,
\beq
\labell{g112}
G_{11} = 16 \pi^{7} l_{P}^{9},
\qquad
G_{4} = \frac{G_{11}}{V_{S^{7}}} = \frac{3}{\pi^{4}L^{7}}G_{11},
\qquad
T_{M5} = \frac{1}{2^{5}\pi^{5}l_{P}^{6}},
\eeq
where $l_{P}$ is the fundamental eleven--dimensional Planck length.
In evaluating the expression
eq.~(\ref{charge2}) we find that all except the last term of eq.~(\ref{3form2}) lead to a vanishing
integral. The following term of $F_{(4)}$ is therefore the only one that is relevant to
the present analysis,
\beq
F_{(4)} =-\frac{L^{2}}{2} \frac{1}{X} d(\mu_{1}^{2}) \wedge d\phi_{1} \wedge
\bar{*}F_{(2)},
\eeq
where
\beq
\left[\bar{*}F_{(2)}\right]_{\alpha_{1}\alpha_{2}} = -\frac{q_{1}}{H^{3/2}} \sin \alpha_{1}.
\eeq
In the end, the total number of giant gravitons $n_{1}$ is found to be
\beq
\labell{number2}
n_{1} = \frac{ q_{1}L^{2}}{4\pi G_{11} T_{M5}} \int d\theta_{1} \, d\phi_{1} \, d\alpha_{1} \, d\alpha_{2}
 \; \cos \theta_{1} \sin \theta_{1} \sin \alpha_{1}.
\eeq
Dropping the $\theta_{1}$ integration (along which the giant gravitons are assumed to be distributed)
from eq.~(\ref{number2})
and integrating over the other angles
we find an expression for the density of giant gravitons as a function $\theta_{1}$,
\beq
\labell{distrib}
\frac{dn_{1}}{d\theta_{1}} = \frac{8 q_{1}N^{1/2}}{\sqrt{2} L} \cos \theta_{1} \sin \theta_{1}.
\eeq
As in section 2, we have detected a distribution of 'electric dipole' sources for $F^{(4)}$ at the
singularity, which is certainly evidence of the presence of giant gravitons.

In order to give further support for this result, we calculate the energy of the above
distribution.
A test--brane analysis showed\cite{toumbas} that a single giant graviton
is associated with an internal momentum given by
\beq
P_{\phi_{1}} = N \sin^{4} \theta_{1}.
\eeq
Combining this result with eq.~(\ref{distrib}), we find the total angular momentum of the distribution,
\beq
\bar{P}_{\phi_{1}} = N \int_{0}^{\pi/2} d\theta_{1} \sin \theta_{1} \frac{dn_{1}}{d\theta_{1}} =
\frac{4\sqrt{2}}{3} \frac{q_{1} N^{3/2}}{L},
\eeq
which corresponds, according to eq.~(\ref{energyGG}), to a total energy
\beq
\labell{totalenergy2}
\bar{E} = \frac{4\sqrt{2}}{3} \frac{q_{1} N^{3/2}}{L^{2}}.
\eeq
The mass of the superstar (\ref{adsmass2}), calculated by conventional supergravity means, is
\beq
\labell{1charge2}
M=\frac{q_{1}}{4G_{4}}.
\eeq
Using eqs.~(\ref{g112}) to write the mass of the superstar in terms of $N$ and $L$, one
shows that it agrees exactly with the total energy eq.~(\ref{totalenergy2})
of a distribution of giant gravitons. This is strong evidence that the asymptotically
$AdS_{4}\times S^{7}$ superstars
are in fact M--theory objects composed of microscopic spherical M5--branes that have expanded on the
seven--sphere.

Following the procedure of the previous section, we can easily generalize to the case of superstars with
none of the four charges vanishing: $q_{i}\neq 0$ ($i=1,2,3,4$). We consider the following embedding
of $S^{7}$ in ${\bf R}^{8}$ with coordinates $x^{1,...,8}$,
\beq
x^{2i-1} = L \mu_{i} \cos \phi_{i}, \;\;\; x^{2i}=L\mu_{i}\sin \phi_{i}.
\eeq
Consequently, a giant graviton moving along $\phi_{i}$ has a radius of
\beq
\rho_{i} = L \sqrt{1-\mu_{i}^{2}}.
\eeq
In analogy with the single charge calculation, the density of gravitons of a certain radius
involves the $F^{(4)}$,
\beq
F^{(4)}_{\rho_{i}\phi_{i}\alpha_{1}\alpha_{2}} = \frac{d\mu_{i}^{2}}{d\rho_{i}}F^{(4)}_{\mu_{i}^{2}
\phi_{i}\alpha_{1}\alpha_{2}} = q_{i}\rho_{i} \sin \alpha_{1}.
\eeq
The corresponding density of giant gravitons for each direction is then
\beq
\frac{dn_{i}}{d\rho_{i}} = \frac{8q_{i}\sqrt{N}\rho_{i}}{\sqrt{2}L^{3}}.
\eeq
Just as above, it is found that the total angular momentum carried by each set of giant gravitons
is
\beq
\bar{P}_{\phi_{i}} = \frac{4\sqrt{2}}{3} \frac{q_{i} N^{3/2}}{L}.
\eeq
These results correspond to the total angular momentum calculated for the superstar solution, and
we have complete agreement between the BPS mass of the superstar (\ref{adsmass2}) and the total energy
of the giant gravitons, $\bar{E}=\sum \bar{P}_{\phi_{i}}/L$.

\section{Giant graviton probes}

In the previous sections, we showed that the two
eleven--dimensional superstar geometries have an interpretation as
the external fields for a collection of giant gravitons. That is,
we should regard these configurations as solutions of the
eleven-dimensional supergravity coupled to the brane actions of
the corresponding giant gravitons. The primary evidence is the
fact that the naked singularities behave as sources for the
appropriate dipoles of the four-form field strength. However, to
complete the analysis, we should also show that we also have a
solution of the brane equations of motion. As a step
in this direction, we consider a giant graviton probe in the
superstar background. We will indeed find expanded configurations
in the limit as $r\rightarrow0$. Further for these configurations,
we will confirm the result: $P_{\phi_1} = N \sin^{p-1} \theta_1$
(where $p+1$ is the dimension of the M-brane's worldvolume). It
was with this result that we showed that the distribution of giant
gravitons derived from flux considerations matched the total
internal momentum of the superstar.

\subsection{Probing superstars in $AdS_7 \times S^4$}

In the following, we present a brief summary of the M2-brane probe
calculations for the superstars presented in section 2. We will
restrict our attention to the simplest case of a singly charged
superstar (\ie only $q_1$ is nonvanishing). The results of the
doubly charged case are less clear --- we will return to this
point in the discussion.

Our giant graviton probe will be a spherical M2-brane inside the
$S^4$ at constant $\theta_1$ and moving on a circle in the
$\phi_1$ direction. We will first consider this configuration at a
finite radius in the anti--de Sitter space (\ie away from the
singularity) and then consider taking the test--brane to $r=0$.
The M2-brane couples to the three form potential $A^{(3)}$
satisfying $dA^{(3)} = F^{(4)}$. Hence to evaluate the worldvolume
action, we need to dualize $*F^{(4)}$ given in eq.~\reef{3form1}
and integrate to find $A^{(3)}$. The result is
\begin{eqnarray}
  F^{(4)} &=& - {1 \over L} H_1 { ( \Delta + 2) \over \Delta^2}  \sin^2 \theta_1
  \cos \theta_1 L d \theta_1  \wedge \sin \theta_2 \cos \theta_2 L d \theta_2 \bigwedge_j
  [Ld \phi_j + A^j_{(1)}] \nonumber \\
& & \qquad \qquad + \ldots \\
&=&  d \left( - H_1 { \sin^3 \theta_1 \over \Delta} \wedge \sin
\theta_2
\cos \theta_2 L d \theta_2  \bigwedge_j [Ld \phi_j + A^j_{(1)}] \right) \nonumber \\
&=& d A^{(3)} .
\end{eqnarray}
The probe action is then
\begin{eqnarray}
  S_2 &=& -T_{M2} \int dt d\theta_2  d\phi_2 \left[  \sqrt{-g} - A^{(6)}_{t \theta_2  \phi_2}
  -  \dot{\phi_1} A^{(4)}_{\phi_1 \theta_2 \phi_2}\right] \nonumber \\
 &=& {N \over L} \int dt  \left[ - \Delta^{-1/2} \sin^2 \theta_1 \left( {f \over H_1}
 - \Delta^{-1}  H_1 \mu_1 ^2 [L \dot{\phi_1} + (H_1^{-1} - 1)]^2 \right)^{1/2} \right.
 \nonumber \\
& & \qquad \left. + {H_1 \over  \Delta} \sin^3 \theta_1 (L
\dot{\phi_1} +H_1^{-1} - 1) \right] ,
\end{eqnarray}

Now fixing the momentum $p \equiv P_{\phi_1}/N$, we find the
Hamiltonian
\begin{equation}
   {\mathcal H} = {N \over L}  \left[\sqrt{ {f \over H_1}  \left({p^2 \over H_1} +
   \tan^2 \theta_1 (p - \sin \theta_1)^2  \right) }  + (1 - H_1^{-1}) p \right].
\end{equation}
Minimizing ${\mathcal H}(r,\theta_1)$ with respect to $\theta_1$
yields $\theta_1 = 0$ or $\sin \theta_1 = p$ for arbitrary values
of $r$. However, to find a true solution of the equations of
motion, we must also minimize with respect to the radius. Setting
$\sin \theta_1 = p$, a short calculation shows that the minimum
energy configuration is at r=0, and that the energy of this
configuration satisfies the BPS relation:
\begin{equation}
   {\mathcal H}(r=0,\,\sin\theta_1=p) ={N\over L} p =   {P_{\phi_1} \over L} .
\end{equation}

\subsection{Probing superstars in $AdS_4 \times S^7$}

The four--form and metric corresponding to the superstars in
$AdS_4 \times S^7$ were given in section 3. The probe which we
shall be studying in this case is a spherical M5-brane inside the
$S^7$ at constant $\theta_1$ and moving on a circle in the
$\phi_1$ direction. Again, we begin by placing the probe away from
the singularity and consider the limit as $r\rightarrow0$. The
M5-brane couples to the six--form potential $A^{(6)}$ satisfying
$dA^{(6)} = *F^{(4)}$ and so our first task is to dualize
$F^{(4)}$ given in eq.~\reef{3form2} and integrate to find
$A^{(6)}$. We find
\begin{eqnarray}
  *F^{(4)} &=& - {2 \over L} H_1 { (2 \Delta + 1) \over \Delta^2}
\sin^5 \theta_1 \cos \theta_1 L d \theta_1  \wedge \sin^3 \theta_2
\cos \theta_2 L
d \theta_2 \nonumber \\
& &  \wedge \sin \theta_3 \cos \theta_3 L d \theta_3 \bigwedge_j
[Ld \phi_j + A^j_{(1)}]
+ \ldots \\
&=&  d \left(- H_1 { \sin^6 \theta_1 \over \Delta} \wedge \sin^3
\theta_2 \cos \theta_2 L d \theta_2  \wedge \sin \theta_3 \cos
\theta_3 L d \theta_3 \bigwedge_j [Ld \phi_j + A^j_{(1)}]
\right) \nonumber \\
&=& d A^{(6)} .
\end{eqnarray}
Evaluating the probe action for the configuration described above
yields
\begin{eqnarray}
  S_5 &=& -T_{M5} \int dt d\theta_2 d\theta_3  d\phi_2 d\phi_3 d\phi_4  \left[  \sqrt{-g}
+ A^{(6)}_{t \theta_2 \theta_3 \phi_2 \phi_3 \phi_4}  +
\dot{\phi_1} A^{(4)}_{\phi_1 \theta_2
\theta_3\phi_2 \phi_3 \phi_4}\right] \nonumber \\
 &=& {N \over L} \int dt  \left[ - \Delta^{-1/2} \sin^5 \theta_1 \left( {f \over H_1}  -
\Delta^{-1}  H_1 \mu_1 ^2 [L \dot{\phi_1} + (H_1^{-1} - 1)]^2 \right)^{1/2} \right. \nonumber \\
 & & \qquad \left. + {H_1 \over  \Delta} \sin^6 \theta_1 (L \dot{\phi_1} +H_1^{-1} - 1)
 \right] .
\end{eqnarray}

Again fixing $p \equiv P_{\phi_1}/N$, we find the Hamiltonian
\begin{equation}
{\mathcal H} = {N \over L}  \left[\sqrt{ {f \over H_1}  \left({p^2
\over H_1} + \tan^2 \theta_1 (p - \sin^4 \theta_1)^2  \right) }  +
(1 - H_1^{-1}) p \right].
\end{equation}
Extremizing ${\mathcal H}$ with respect to $\theta_1$ yields
minima at $\theta_1 = 0$ and $\sin^4 \theta_1 = p$ independent of
the radius. The $r$-dependence of the Hamiltonian is essentially
as before and we once again find a solution to the equations of
motion at $r=0$ satisfying the BPS relation :
\begin{equation}
   {\mathcal H}(r=0,\,\sin^4 \theta_1 = p) =   {P_{\phi_1} \over L} .
\end{equation}

\section{Discussion}

In this note, we have extended the analysis of ref.~\cite{myers1}
to the context of M--theory. In particular, we have studied
superstar solutions of eleven--dimensional supergravity in
asymptotically $AdS_7\times S^4$ and $AdS_4\times S^7$
space--times. While these BPS configurations contain naked
singularities, we have argued that M--theory provides a natural
external source in the form of an ensemble of giant gravitons.
Hence we should regard the superstars as solutions of the combined
equations of motion of eleven--dimensional supergravity coupled to
the worldvolume actions of the corresponding giant gravitons.

In section 4, we considered the dynamics of giant gravitons in the superstar
backgrounds. With this analysis, we were able to
recover the result that the expanded branes satisfy:
\beq
\labell{expand}
P_{\phi_1} = N \sin^{p-1} \theta_1
\eeq
where $p+1$ is the dimension of the M-brane's worldvolume. We have also
confirmed that this result applies for a giant graviton probe in
the type IIB superstar background considered in
ref.~\cite{myers1}. It is curious that these relations between the
internal momentum of the various giant gravitons and their
expansion in the internal space is identical to that derived in
the corresponding $AdS_m\times S^n$ background. Certainly the
energy--momentum relations for the BPS configurations must hold in
the superstar backgrounds, as this is dictated by supersymmetry.
However, the latter does not require the dynamical details such as
the size to remain unperturbed in the superstar backgrounds. It
would be interesting to investigate whether this result is simply
a coincidence or whether there is a deeper reason why the
expansion of the giant gravitons matches in these different
backgrounds.

One should also question the validity of the result presented in
eq.~\reef{expand} since it relies on pushing past the expected
regime of validity of the test--brane analysis. We expect, \eg, that the
usual Born--Infeld form of the action is only a leading order
result that should be valid to describe the low energy dynamics of
branes in slowly varying background fields. We must expect that
there are higher derivative corrections that must become important
in rapidly varying backgrounds, \eg, in regions of strong
curvature. For instance the leading order corrections to the worldvolume action
can be systematically calculated in superstring theory
\cite{green}. In the present analysis, we begin by placing the
giant gravitons away from the origin but then find that minimizing
the energy requires that we set $r=0$. However, this is the
position of the naked singularity in the singly charged superstar,
and so in fact, it is a location where the curvatures are
diverging. The fact that we get a simple and consistent result
seems to indicate that the higher curvature corrections must
cancel out in the final analysis. However, clearly this point
deserves a better understanding. In particular, one should note that
the analysis in section 4 only considers the case of a singly charged
superstar. We were unable to confirm the same results for the general
case, but it may well be that strong curvature effects play an important
role in these cases. Certainly, the analysis for the type IIB superstars
\cite{myers1} show the multiply charged systems to be more exotic.

\section*{Acknowledgements}
Research by FL and RCM is supported in part by NSERC of Canada and
Fonds FCAR du Qu\'ebec. DCP is funded by the Engineering and
Physical Sciences Research Council of Britain. FL would like to thank the
Perimeter Institute for Theoretical Physics and the University of
Waterloo's Department of Physics for their kind hospitality during
this work. We would like to thank Clifford Johnson for useful
conversations and correspondence. DCP would also like to thank Ken
Lovis and especially Douglas Smith for extensive conversations. While this paper
was being completed, ref.~\cite{vijay} appeared which overlaps
with the material presented in sections 2 and 3.

\appendix

\section{Dual giant graviton probes}

We have also considered probing the M--theory superstars with
dual giant graviton probes, in analogy with the analysis presented
in ref.~\cite{myers1}.

\subsection{Dual giants in $AdS_7 \times S^4$}

In $AdS_7 \times S^4$, the dual giants are  spherical M5-branes, spanning the $S^5$ of the AdS$_7$ space at constant r. They orbit on the $S^4$ at constant $\theta_1$ and $\theta_2$  with fixed angular momentum $P_{\phi_i}$ conjugate to the angles $\phi_i$.

The probe Lagrangian takes the following form (after integrating over the 5-sphere):
\begin{eqnarray}
  {\mathcal L} &=& {\tilde{N} \over  L} \left[ - \Delta^{1/2} \left({r \over \tilde{L}}\right)^5 \left( f \left( \mu_0 ^2 + \sum_i {\mu_i ^2 \over H_i}\right) - \sum_i H_i \mu_i ^2 [L \dot{\phi_i} + (H_i^{-1} - 1)]^2 \right)^{1/2} \right. \nonumber \\
 & & \qquad \left. + {r^6 \over \tilde{L}^6} \Delta + \sum_i {\mu_i ^2  q_i \over \tilde{L}^4} (L \dot{\phi_i} - 1) \right]
\label{Probe3}
\end{eqnarray}

Here  $\tilde{L}=2L$ is the radius of the AdS space and $\tilde{N} = L \tilde{L}^2 A_2 T_2$, where $A_2$ is the area of a unit $S^2$ and $T_2$ is the tension of an M2 brane. Fixing the value of $\tilde{p_i} \equiv P_{\phi_i}/\tilde{N}$ we find the Hamiltonian:
\begin{eqnarray}
{\mathcal H} &=& {\tilde{N} \over L} \left[\sqrt{ f \left( \mu_0 ^2 +  \sum_i {\mu_i ^2 \over H_i} \right)  \left( \sum_i {(\tilde{p}_i -  {\mu_i ^2 q_i \over \tilde{L}^4})^2 \over  H_i \mu_i ^2} + \Delta {r^{10} \over \tilde{L}^{10}} \right) } \right. \nonumber \\
 & & \qquad \left. + \sum_i (1 - H_i^{-1}) \left(\tilde{p}_i -  {\mu_i ^2 q_i \over \tilde{L}^4} \right) \, - \Delta {r^6 \over \tilde{L}^6} + \sum_i {\mu_i ^2 q_i \over \tilde{L}^4} \right]
\end{eqnarray}

Mininima of this action saturating the BPS bound ${\mathcal H} = (N/L) \sum p_i$ occur whenever the following equations are satisfied:

\begin{equation}
{r^4 \over \tilde{L^4}} \mu_i^2 = \tilde{p}_i - {q_i \mu_i^2 \over \tilde{L^4}}.
\end{equation}

Using the constraint that $\sum \mu_i^2=1$, one can reduce these equations to
the same form as presented in eqs.~(28) and (29) of ref.~\cite{myers1}.

\subsection{Dual giants in $AdS_4 \times S^7$}

In $AdS_4 \times S^7$, the giant gravitons are M5-branes whereas the dual giants are  spherical M2-branes, spanning the $S^2$ of the AdS$_4$ space at constant r. They orbit on the $S^7$ at constant $\theta_1$, $\theta_2$ and $\theta_3$ with fixed angular momentum $P_{\phi_i}$ conjugate to the angles $\phi_i$.

The probe Lagrangian takes the following form (after integrating over the 2-sphere):
\begin{eqnarray}
{\mathcal L} &=& {\tilde{N} \over L} \left[ - \Delta^{1/2} {r^2 \over \tilde{L}^2} \left( \sum_i {\mu_i ^2 \over H_i} f - \sum_i H_i \mu_i ^2 [L \dot{\phi_i} + (H_i^{-1} - 1)]^2 \right)^{1/2} \right. \nonumber \\
 & & \qquad \left. + {r^3 \over \tilde{L}^3} \Delta + \sum_i {\mu_i ^2  q_i \over \tilde{L}} (L \dot{\phi_i} - 1) \right]
\label{Probe2}
\end{eqnarray}

Here we have introduced $\tilde{L}=L/2$ as the radius of the AdS space and $\tilde{N} = L \tilde{L}^5 A_5 T_{5}$ where $A_5$ is the area of a unit $S^5$ and $T_5$ is the tension of an M5-brane. Fixing the value of $\tilde{p_i} \equiv P_{\phi_i}/\tilde{N}$ we find the Hamiltonian:
\begin{eqnarray}
{\mathcal H} &=& {\tilde{N} \over L} \left[\sqrt{ \left( \sum_i {\mu_i ^2 \over H_i} f \right)  \left( \sum_i {(\tilde{p}_i -  {\mu_i ^2 q_i \over \tilde{L}})^2 \over  H_i \mu_i ^2} + \Delta {r^4 \over \tilde{L}^4} \right) } \right. \nonumber \\
 & & \qquad \left. + \sum_i (1 - H_i^{-1}) \left(\tilde{p}_i -  {\mu_i ^2 q_i \over \tilde{L}} \right) \, - \Delta {r^3 \over \tilde{L}^3} + \sum_i {\mu_i ^2 q_i \over \tilde{L}} \right]
\end{eqnarray}

Mininima of this action saturating the BPS bound ${\mathcal H} = (N/L) \sum p_i$ occur whenever the following equations are satisfied:

\begin{equation}
{r \over \tilde{L}} \mu_i^2 = \tilde{p}_i - {q_i \mu_i^2 \over \tilde{L}}.
\end{equation}

\bibliographystyle{plain}

\end{document}